\documentclass[12pt,twocolumn]{article}
\usepackage{graphicx}
\usepackage{color}
\usepackage{hyperref}
\usepackage{caption}
\usepackage{setspace}
\usepackage[margin=1in]{geometry}
\usepackage{ulem}
\usepackage[T1]{fontenc}
\usepackage{amsmath}
\usepackage[utf8]{inputenc}
\usepackage{hyperref}
\usepackage{epstopdf}
\onecolumn
\begin{document}
\begin{center}
{\Large {\bf Mixed phase and Tetracritical Behaviour of Dilute 3D Heisenberg Magnet}}
\end{center}
\vskip 0.5cm
\begin{center}
{\it Nepal Banerjee}\\
{\it Department of Physics,University of Seoul,South Korea}
\vskip 0.2 cm
{\bf {Email}}: nb.uos1989@gmail.com
\end{center}
\vskip 0.5cm

%%%%%%%%%%%%%%%%%%%%%%%%%%%%%%%%%%%%%%%%%%%%%%%%%%%%%%%%%%%%%%%%%%%%%%%%%%%%%%%%%
%                            Abstract                                           %
%%%%%%%%%%%%%%%%%%%%%%%%%%%%%%%%%%%%%%%%%%%%%%%%%%%%%%%%%%%%%%%%%%%%%%%%%%%%%%%%%
\begin{abstract}
Recent breakthrough discovery of twisted bilayer $CrI_3$ system,where both ferromagnetic and antiferromagnetic order coexist act as a main motivation for simulating simple magnetic system where this type of mixed phase arises \cite{cri3}.Here we have simulated the dilute magnetic alloy with generic type {$\bf {A_p B_{1-p}}$}.Here \textbf{A} and \textbf{B } are ferromagnetic  and antiferromagnetic  type magnetic atoms.Here we have studied the mixed phase and tetracritical behaviour of that dilute magnet.We have simulated the critical behaviour of that kind of mixed phase at different doping strength(p).Here we have observed mixed phase rather than spin-glass phase ({$\bf{SG}$}) in this site-random({\bf{SR}}) disorder model.Here we have used classical Monte-Carlo simulation with Heisenberg spin and use 3D simple cubic lattice for this simulation.
\end{abstract}

%%%%%%%%%%%%%%%%%%%%%%%%%%%%%%%%%%%%%%%%%%%%%%%%%%%%%%%%%%%%%%%%%%%%%%%%%%%%%%%%%%
%%%%%%%%%%%%%%%%%%%%%%%%%%%%%%%%%%%%%%%%%%%%%%%%%%%%%%%%%%%%%%%%%%%%%%%%%%%%%%%%%%
%                         Introduction                                           %
%%%%%%%%%%%%%%%%%%%%%%%%%%%%%%%%%%%%%%%%%%%%%%%%%%%%%%%%%%%%%%%%%%%%%%%%%%%%%%%%%%
%%%%%%%%%%%%%%%%%%%%%%%%%%%%%%%%%%%%%%%%%%%%%%%%%%%%%%%%%%%%%%%%%%%%%%%%%%%%%%%%%%
\section{Introduction:}
In recent year the discovery of several interesting van der Waals(\textbf{vdW}) magnetic material introduced to us a fascinating world of strongly correlated phase of matter\cite{burch2018magnetism,gong2017discovery,sinova,park2016opportunities,jiang2021recent,chit2,cri3_1}.The highly complex magnetic phase of those van der Waals(\textbf{vdW}) material is always a challenge for theoretical condensed matter physics and several interesting model has been introduced to explain those complex phase\cite{TD,yogendra,she2010stability,bistritzer,jeil,sumilan}.Disorder plays an important roles and add more flavour on those correlated phase\cite{rhodes2019disorder,vojta2019disorder,edwards,SK,parisi,bhatt}.In fact real material are full of imperfection,defect and impurity so it is essential to consider disorder and that reveal the realistic scenario and exact phase of those magnetic material.Different complex magnetic phase usually appear in presence of different type of magnetic atom, which are interacting ferromagnetically or antiferromagnetically between each other.Experimentally this type of  materials are quite available which contain different type of magnetic and non-magnetic atoms in a different ratio and their mutual interaction generically emerges as a complex magnetic phase.In presence of ferromagnetic and antiferromagnetic exchange interaction and from their competitive interaction several type of frustrated phases,spin glass(SG) phases emerges from those materials\cite{mezard1987spin,ryan1992recent,Dchoudhary,chandan}.There exist another class of interesting magnetic phase,which we call as mixed phase where both FM and AFM phase coexist with each other and that kind of phase is emerging from a class of van der Waals(vdW) magnetic system with generic type {{$\bf A_x B_{1-x}$}}.Here pure \textbf{A} has ferromagnetic ordering and \textbf{B} has antiferromagnetic ordering.As an example we can mention number of interesting magnetic material where we can observe this type of mixed magnetic phase.Examples are $\bf {Fe(Pd_xPt_{1-x})_3}$, where $\bf{FePd_3}$ is ferromagnetic and $\bf {FePt_3}$ is antiferromagnetic.$\bf {(Mn_pFe_{1-p})WO_3}$ where $\bf {MnWO_3}$ and $\bf{FeWO_3}$ has different type of antiferromagnetic ordering.$\bf{UAs_pSe_{1-p}}$,where {\bf{USe}} order ferromagnetically and {\bf{UAs}} order antiferromagnetically.Also the materials like {$\bf {Eu_xSr_{1-x}S}$} and {$\bf{Au_{1-x}Fe_x}$} usually host mixed phase,where both ferromagnetic and spin glass(SG) phase co-exist.Similarly there are several type of magnetic,non-magnetic,high-Tc superconducting materials like \textbf{CeRhIn5}, \textbf{Rh-doped CeIrIn5}, \textbf{UGe2}, where we have observed this type of mixed phase and tetracritical behaviour in a intermediate doping(p) during quantum phase transition(\textbf{QPT})\cite{sachdev,sachdev2,sachdev3}.We can describe this mixed magnetic phase with Landau-Ginzburg like phenomenological theory and that gives us a very clear and qualitative idea about that mixed phase and their transition properties.Here we have given a pedagogical overview of that mixed phase based on that Landau-Ginzburg theory.Here the free energy(F) of that system in presence of FM and AFM magnetic ordering is described by

%%%%%%%%%   Fig:1 %%%%%%%%%%%%%%%%%%%%%%%%

\begin{figure}[htp]
\centering
\includegraphics[scale=0.30]{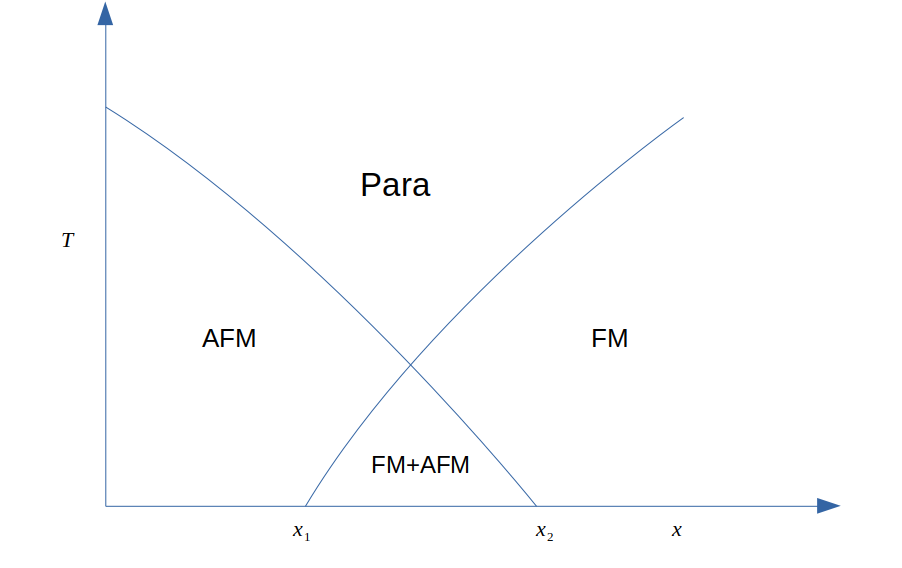}
\caption{Here we have presented schematically the phase diagram and phase transition for two competitive order parameter.Here two second order phase transition meet at tetracritical point and we can observe mixed phase at the intermediate value of doping.Here ferro(FM) and antiferro(AFM) is two competitive order parameter.}
\label{}
\end{figure}

%%%%%%%%%%%%%%%%%%%%%%%%%%%%%%%%%%%%%%%%%%

\begin{eqnarray}
F &=& F_M+ F_\psi + F_{int}\\
F_{\psi} &=&\frac{1}{2}(\nabla \psi)^2-\alpha {\psi}^2 + u/2  \ {\psi}^4 \\
F_{M}&=&\frac{1}{2}(\nabla M)^2-\alpha_M M^2 + v/2 \ M^4 \\
F_{int}&=& w \ {{\psi}^2{M}^2}
\end{eqnarray}
Here ${\bf F_M}$ and ${\bf F_{\psi}}$ is the free energy for AFM and FM phase.Here ${\bf {F_{int}}}$ indicate the interaction between the two phase.Here $\bf {\alpha=a(x-x_1),\alpha_M=a(x_2-x)} $.Now for observing the stable phase of matter we have to minimize the free energy with respect to different order parameter of that system.So the condition for that minimization are following 
\begin{eqnarray}
\frac {\partial F}{\partial \psi}&=&0,\\
\frac {\partial F} {\partial M} &=& 0,
\end{eqnarray}
With this minimization condition we will get different combination of $\bf{\psi^*, M^*}$ for which free energy will show minima for a specific condition and that is going to determine the stable phase of that system at that specific condition.With this minimization technique we will get four combination of ($\bf{\psi^*,M^*}$) and those values are following    
\begin{eqnarray*}
(|\psi|,|M|)&=& (0,0),(0,\sqrt{\alpha_{M} /v}),(\sqrt{\alpha/u},0),(\psi^*,M^*).
\end{eqnarray*}
\begin{eqnarray}
\alpha \psi*^2 &=& \frac{w'-v'}{w'^2- u'v'} \\
\alpha_M M*^2 &=& \frac{w'- u'}{w'^2 -u'v'} 
\end{eqnarray}  
\begin{eqnarray*}
w'= w/\alpha \alpha_M , u' = u/\alpha^2,v' =v/\alpha^2_M
\end{eqnarray*}
Here (0,0),(0,$\sqrt{\alpha_{M} /v }$) solution indicate the para phase and pure AFM phase.Here ($\sqrt{\alpha/u}$,0) indicate the pure FM phase and last solution ($\psi^* ,M^*$)indicate mixed phase.Here four phase is coexising at tetracritical point and we can also think in another way that this multicritical point seperating four phase of this system.Here the fourth solution where both order parameter $\psi^*$ and $M^*$ is non-zero indicating that there exist a stable phase where two competitive order $\psi$ and M can coexist and that is called mixed phase.So near at $x=x_1$,
$\psi^*$ is almost zero and $ M^*\neq 0 $.Now after putting  $\psi^*=0$ at the Eq(7) we will get the value of x 
\begin{eqnarray}
\frac{w'-v'}{w'^2- u'v'}=0\\
w'=v'\\
w/\alpha\alpha_M= v/\alpha^2_M\\
\alpha_M w= v \alpha \\
a_M(x_2- x)w = v a(x-x_1)\\
x= \frac{w a_M x_2+ av x_1}{a v+ a_M w} 
\end{eqnarray}
Similarly near $x=x_2$ it can be shown from same argument that near at $x_2$,M is almost zero and $\psi^* \neq 0$. If we put zero at Eq(8) we will get 
\begin{eqnarray}
\frac{w'- u'}{w'^2 -u'v'}=0\\
w'=u'\\
w/\alpha \alpha_M = u/\alpha^2\\
w\alpha = u\alpha_M\\
w a(x-x_1)= u a_M(x_2-x) \\
x= \frac{a_M x_2 v + w a x_1}{w a+a_M u}
\end{eqnarray} 
So for existing mixed phase the tuning parameter(x) which is here doping(p) must be with in  this interval. 
\begin{eqnarray}
\frac{w a_M x_2+ av x_1}{a v+ a_M w}<x<\frac{a_M x_2 v + w a x_1}{w a+a_M u}
\end{eqnarray}
For getting the idea of its width we have to substract two extreme limit of x,which are near at $x_1$ and $x_2$ respectively.\\
\begin{eqnarray}
W &=&\frac{a_M x_2 v + w a x_1}{w a+a_M u} - \frac{w a_M x_2+ av x_1}{a v+ a_M w}\\
 &=& a a_M(x_2 -x_1) (uv - w^2) \\
 &=&W_0( uv -w^2) 
\end{eqnarray}
Here ${W_0=a a_M(x_2 -x_1)}$ is a positive constant.
We can now easily explain the tetracritical and bicritical behaviour from this expression of width of tuning parameter(x) where two phase coexist.Here  
\begin{eqnarray}
W=W_0( uv -w^2)
\end{eqnarray}
With this simple analysis we can see that the width of tuning parameter(x) where mixed phase exist has finite positive value when $\bf {uv> w^2}$.Here system is showing a {\bf tetracritical} point and behaviour,where two continuous second order phase transition intersect.Here interaction strength {\bf{w}} is weak.When $\bf{uv=w^2}$ then the width of mixed phase will vanish and multicritical point transform into {\bf bicritical} point and system will show bicritical behaviour.Here system goes through a first order phase transition.Several theoretical approach based on Renormalization group theory is performed by Kosterlitz,Nelson,Fisher,Aharony and Mukamel where they have presented the exact nature of phase transition and properties of bicritical,tetracritical like multi-critical phase \cite{kosterlitz,mukamel,aharony,aharony3,fisher1974spin,aharony2,pm}.Here we have attempted to probe the microscopic behaviour of phase transition of those kind of dilute magnetic alloy,where both ferro and antiferro like atom present with simple classical Monte-Carlo simulation.Here we have simulated a mixed magnetic alloy which have  generic formula {$\bf{A_p B_{1-p}}$}.Here \textbf{A} is ferromagnetic type of atom and \textbf{B} is antiferromagnetic atom.Here we use 3D Heisenberg model with cubic lattice.We are using the model which is originally proposed by Matsubara in their celebrated paper\cite{matsubara1996mixed}.Here we basically using site random disorder where we are randomly filling \textbf{B} type antiferromagnetic atom with a specific filling fraction({\bf{1-p}}) and they are interacting with each other with antiferromagnetic exchange interaction.This type of site random disorder is easily accessible by experiment and order state is more favourable in this type of disorder comparative to bond random disorder(BR).Here we have observed a mixed phase in intermediate doping of {\bf {B}} atom and we have presented the critical phenomena of that system in different strength of doping(p) of \textbf{B} atom.Here we have able to probe the mixed phase behaviour and able to simulate tetracritical behaviour of a generic class of dilute magnetic alloy with ab-initio classical Monte-Carlo simulation.We have presented the spin configuration during the transition at different T and that gives us a very clear idea of those interesting magnetic phase where tetracritical phase appears.So we have planed our paper in a following way.First we are going to describe the model Hamiltonian and then explain the simulation method and results of our simulation for different strength of doping(p).In last section we have discussed our all results and make a conclusion.

%%%%%%%%%%%%%%%%%%%%%%%%%%%%%%%%%%%%%%%%%%%%%%%%%%%%%%%%%%%%%%%%%%%%%%%%%%%%%%%%
%                Model Hamiltonian                                             %
%                                                                              %
%%%%%%%%%%%%%%%%%%%%%%%%%%%%%%%%%%%%%%%%%%%%%%%%%%%%%%%%%%%%%%%%%%%%%%%%%%%%%%%%
\section{Model Hamiltonian:}
Here we are describing a model Hamiltonian where two type of magnetic ion A and B which are distributed randomly at the site of 3D simple cubic lattice with concentration of p and (1-p).The exchange interaction between A-A ,B-B and A-B (B-A) is +J ,-J ,+J respectively.We can write the Hamiltonian as 
\begin{eqnarray}
H=-\frac{J}{2}\sum_{<i,j>} [1+(\epsilon_i +\epsilon_j)-\epsilon_i \epsilon_j] \vec S_i .\vec S_j
\end{eqnarray}
Here $\vec S_i=(S_i^x,S_i^y ,S_i^z)$ represent classical three component spin with unit magnitude sitting at each lattice point i.Here $\epsilon_i=+1$ for A atom and $\epsilon_i=-1$ for B atom.Hamiltonian has symmetry with respect to p and (1-p) and phase diagram is symmetry about p=0.5.In this model we observe FM ,AFM and mixed phase where FM and AFM order coexist. 

%%%%%%%%%%%%%%%%%%%%%%%%%%%%%%%%%%%%%%%%%%%%%%%%%%%%%%%%%%%%%%%%%%%%%%%%%%%%%%%%%
%                                                                               %
%                   Simulation Methodology and Results                          %
%                                                                               %
%%%%%%%%%%%%%%%%%%%%%%%%%%%%%%%%%%%%%%%%%%%%%%%%%%%%%%%%%%%%%%%%%%%%%%%%%%%%%%%%%
\section{Simulation Methodology and Results:}
Here we have used the classical Monte-Carlo simulation technique for simulation of disorder induced 3D heisenberg magnet.Here we have simulated {$\bf A_p B_{1-p}$} type disorder heisenberg magnet where \textbf{A} type atom is filled randomly at different site of that simple cubic lattice with the filling factor p and \textbf{B} type atom is filled with a filling factor(1-p).Here \textbf{A} type atom interacting each other with ferromagnetic exchange interaction and \textbf{B} type atom interacting with each other with anti-ferromagnetic interaction and A and B interact with
%%%%%%%%%%%%%% Fig:2 %%%%%%%%%%%%%%%%%%%%%%%%%
\begin{figure}[htp]
\centering
\includegraphics[scale=0.30]{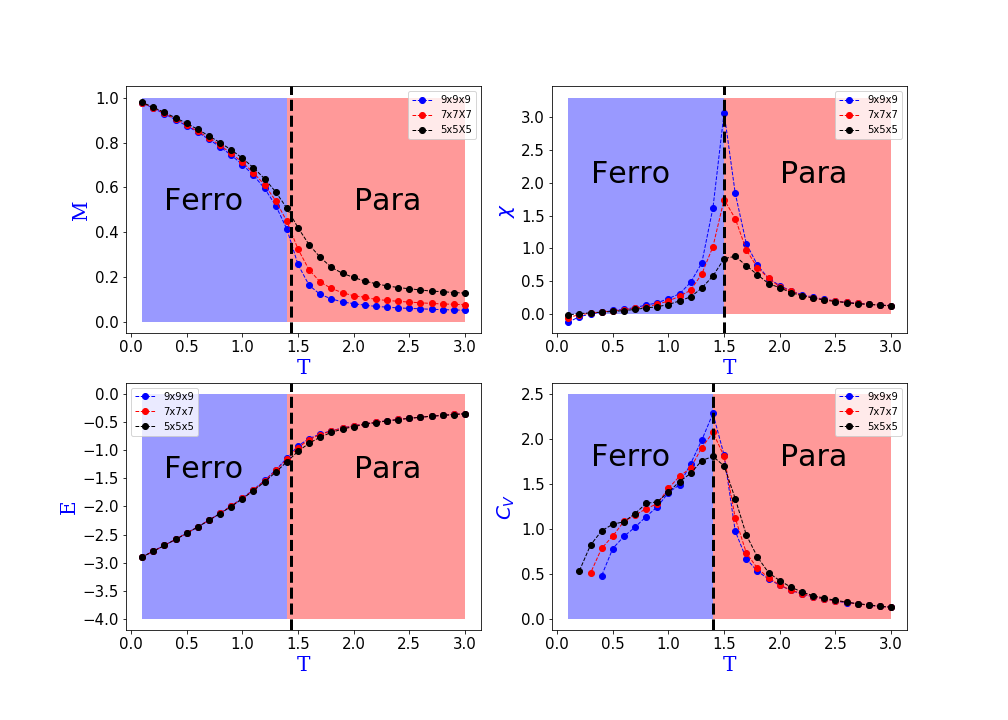}
\caption{Here we have presented the critical behaviour for p=1.0.Here we have presented M,{\large{$\chi$}},E,$C_v$ with T.Here we have presented the critical behaviour of {$\bf {A_pB_{1-p}}$} magnetic alloy.Here A is ferromagnetic atom and B is antiferromagnetic atom.}
\label{}
\end{figure}
%%%%%%%%%%%%%%%%%%%%%%%%%%%%%%%%%%%%%%%%%%%%%%
ferromagnetic exchange interaction.Here we use single spin flipping Metropolis algorithm for perform the simulation\cite{landau1981phase,banerjee2023simulation,banerjee2023critical,pinettes1998phase,DHR,muktish,olivia}.Here we consider the magnitude of A type spin as $|S_A|=1.0$ and $|S_B|=1.01$.Here the $J_x ^A=J_y ^A=J_z^A=+J=1$ and $J_x ^B=J_y ^B=J_z^B=-J=-1$. Here we have used inbuilt uniform random number generator of our computer for filling the lattice site with specific filling factor.If our random number value $r \leq p$ then we filled with A type of atom otherwise we have filled with B type of atom.Here we have used the following formula for simulating the different component of A-type ferromagnetic classical spin.
%%%%%%%%%%%%%%%%%%%%%%%%%%%%%%%%%%%%%%%%%%%%%%%
\begin{eqnarray}
S_x ^A &=&|S_A|\sin(\theta) \cos(\phi)\\
S_y ^A &=&|S_A|\sin(\theta) \sin(\phi)\\
S_z ^A &=&|S_A|\cos(\theta)  
\end{eqnarray} 
Similarly we have simulated the B-type antiferromagnetic classical spin with the following formula
\begin{eqnarray}
S_x ^B &=&|S_B|\sin(\theta) \cos(\phi)\\
S_y ^B &=&|S_B|\sin(\theta) \sin(\phi)\\
S_z ^B &=&|S_B|\cos(\theta)  
\end{eqnarray} 
Here we have use the following formula for generate the random value of $\theta$ and $\phi$ as $\theta=cos^{-1}(2u_1-1)-\pi$ and $\phi=2 \pi(1-u_2)$.Here $u_1$ and $u_2$ is computer generated random number.
This all A and B atoms are randomly sitting on 3D lattice grid with (i,j,k) coordinate.
Here we have used $5 \times 10^4 $ MC steps for equilibrate the system and another $5 \times 10^4 $ MC steps for calculate the average of different thermodynamic variables at different T.
%%%%%%%%%%%%%%% Fig:3 %%%%%%%%%%%%%%%%%%%%%%%%%
\begin{figure}[htp]
\centering
\includegraphics[scale=0.20]{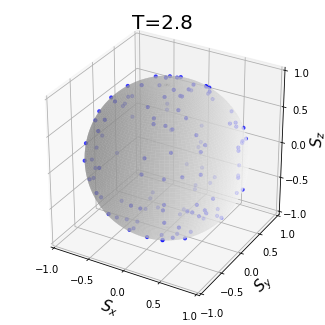}
\includegraphics[scale=0.20]{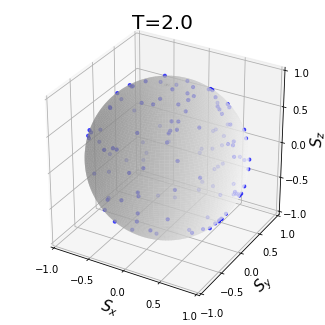}
\includegraphics[scale=0.20]{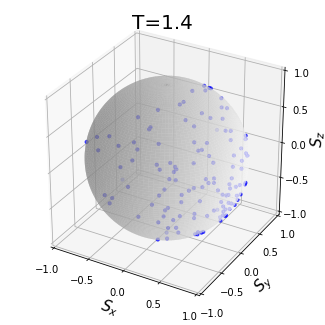}
\includegraphics[scale=0.20]{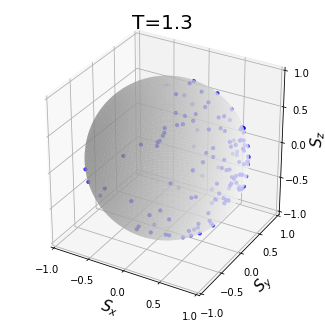}
\includegraphics[scale=0.20]{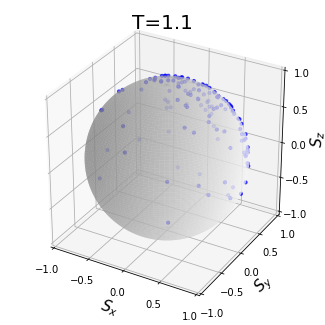}\\
\includegraphics[scale=0.20]{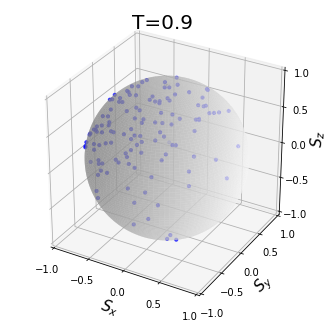}
\includegraphics[scale=0.20]{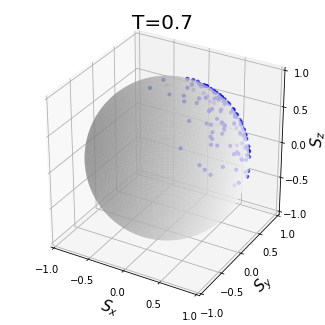}
\includegraphics[scale=0.20]{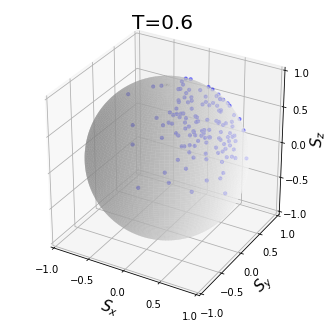}
\includegraphics[scale=0.20]{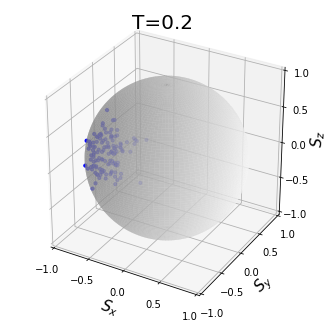}
\includegraphics[scale=0.20]{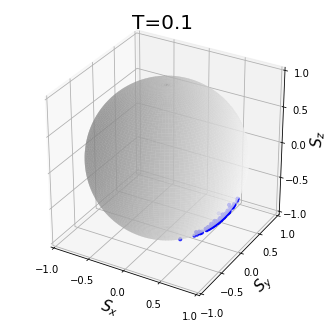}
\caption{Here we are observing the  spin configuration at different T for p=1.0.Here we are using lattice size $5 \times 5 \times 5$.}
\label{}
\end{figure}
Here we have considered $5 \times 5 \times 5$, $7 \times 7 \times 7$ and  $9 \times 9 \times 9 $ lattice size for simulation at different value of p.Here we have studied the critical behaviour of the disorder system at different value of p.Where value of p is p=1.0,\ 0.9,\ 0.8,\ 0.7,\ 0.6,\ 0.5.Here we have presented critical behaviour for each value of p.Here we are using the following formula for calculate the different thermodynamic variables
\begin{eqnarray}
M &=&\sqrt{m_x^2 +m_y^2 +m_z^2}\\
C_{V}&=& L^3(<E^2>-<E>^2)\\
\chi&=& L^3(<M^2>-<M>^2)
\end{eqnarray} 
Here $m_x=\sum S_x/L^3$,$m_y=\sum S_y/L^3$ and $m_z=\sum S_z /L^3$. 
%%%%%%%%%%%%%%%%%%%%%%%%%%%%%%%%%%%%%%%%%%%%%%%%
Here we have first simulated the critical phenomena for p=1.0.In this specific value of p there is only A type of atom is present and they interacting ferromagnetically.So we just observed the critical behaviour of clean 3D heisenberg ferromagnet where the value of $T_c=1.44(29)$.We also observed the same type of behaviour in this case from our simulation.Here the saturation value of spontaneous magnetization(M) exactly reached 1 at low temperature which we can see from Fig(2-3).Here 
%%%%%%%%%%%%%%%% Fig:4 %%%%%%%%%%%%%%%%%%%%%%%%%
\begin{figure}[htp]
\centering
\includegraphics[scale=0.30]{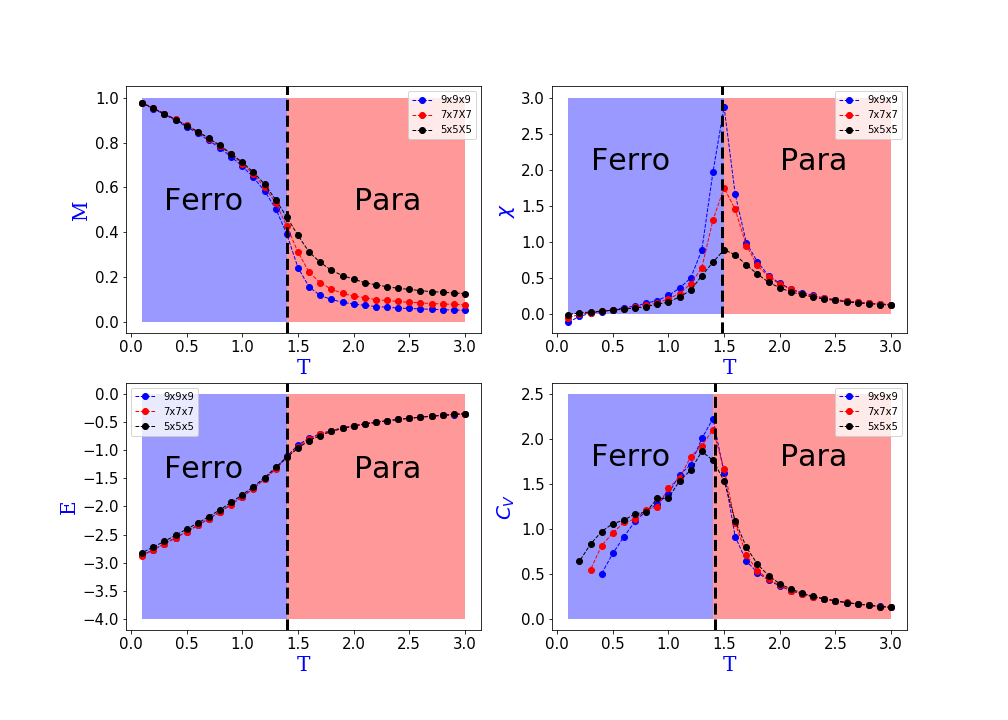}
\caption{Here we have presented the critical behaviour for p=0.90.Here we have presented M,{\large{$\chi$}},E,$C_v$ with T.Here we have presented the critical behaviour of {$\bf {A_pB_{1-p}}$} magnetic alloy.Here A is ferromagnetic atom and B is antiferromagnetic atom.}
\label{}
\end{figure}
%%%%%%%%%%%%%%% Fig:5 %%%%%%%%%%%%%%%%%%%%%%%%%%%
\begin{figure}[htp]
\centering
\includegraphics[scale=0.20]{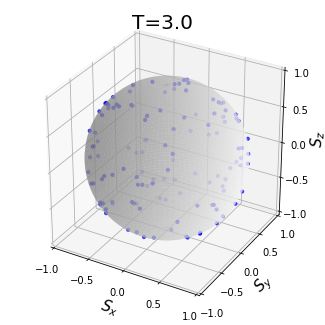}
\includegraphics[scale=0.20]{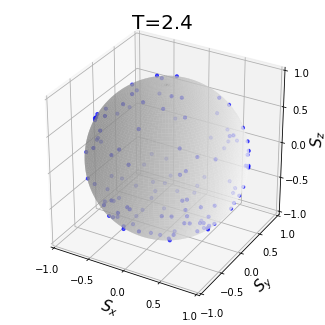}
\includegraphics[scale=0.20]{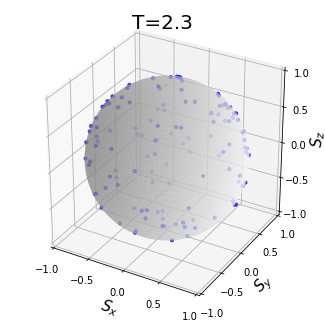}
\includegraphics[scale=0.20]{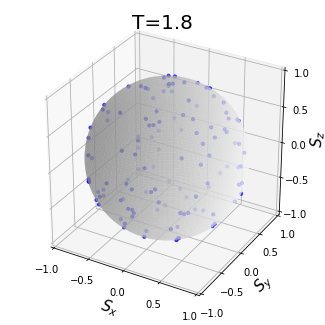}
\includegraphics[scale=0.20]{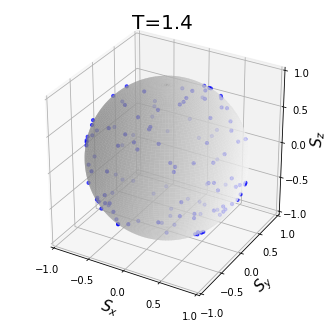}\\
\includegraphics[scale=0.20]{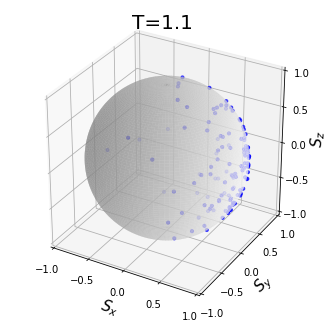}
\includegraphics[scale=0.20]{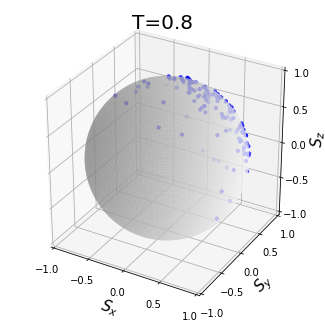}
\includegraphics[scale=0.20]{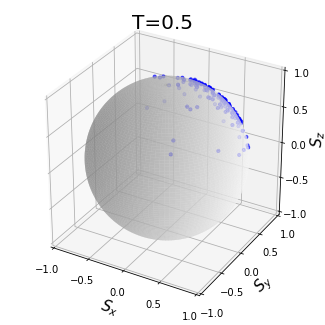}
\includegraphics[scale=0.20]{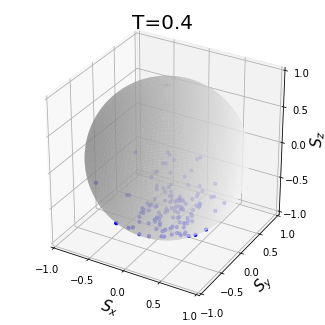}
\includegraphics[scale=0.20]{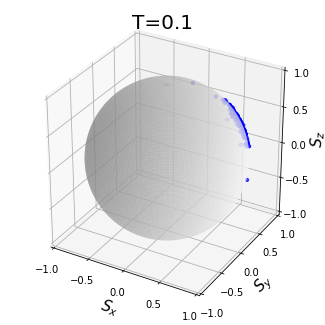}
\caption{Here we are observing the spin configuration at different T for p=0.9.Here we have used $5 \times 5 \times 5$ lattice size.}
\label{}
\end{figure}
%%%%%%%%%%%%%%%%%%%%%%%%%%%%%%%%%%%%%%%%%%%%%%%%%
we have also study the behaviour of $\chi$ and $C_v$ during the transition for different lattice size $5 \times 5 \times 5 $,$7 \times 7 \times 7 $,$9 \times 9 \times 9 $.Here we clearly observed that the peak of $\chi$ and $C_v$ is showing singular behaviour and diverging with system size(L).Here we observe that the system  manifest the critical behaviour of pure 3D Heisenberg system and system showing perfectly ferromagnetic ground state.This results perfectly prove that our simulation is perfectly simulating the critical behaviour of {$\bf A_pB_{1-p}$} type of magnetic alloy and its critical behaviour.  
%%%%%%%%%%%%%%%%%% Fig:6 %%%%%%%%%%%%%%%%%%%%%%%%%  
\begin{figure}[htp]
\centering
\includegraphics[scale=0.30]{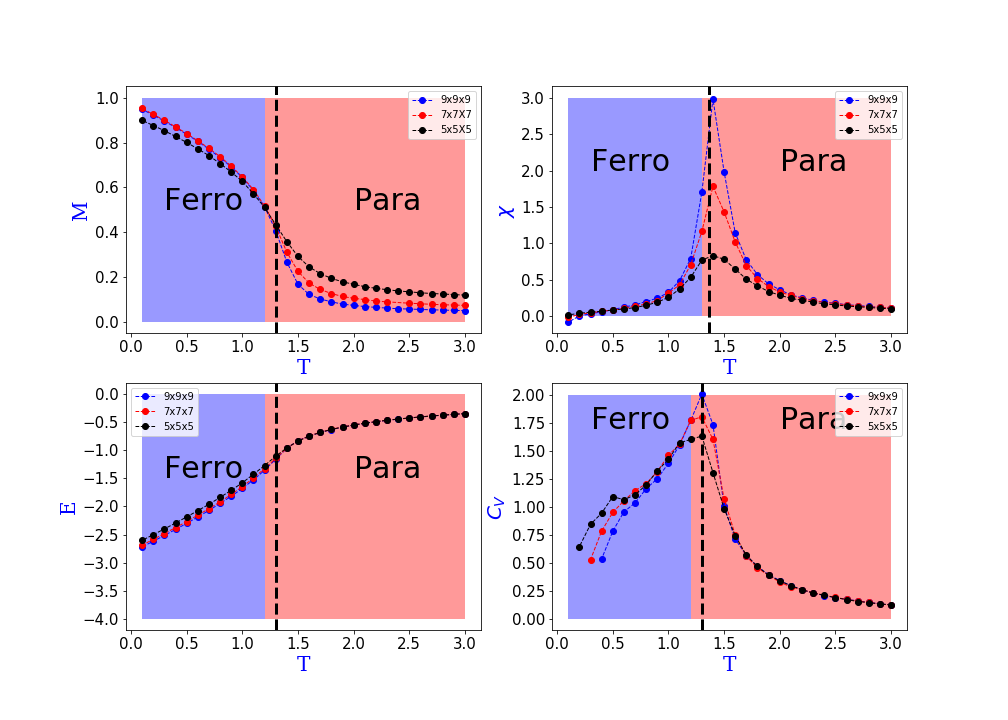}
\caption{Here we have presented the critical behaviour for p=0.80.Here we have presented M,{\large{$\chi$}},E,$C_v$ with T.Here we have presented the critical behaviour of {$\bf{A_pB_{1-p}}$} magnetic alloy.Here A is ferromagnetic atom and B is antiferromagnetic atom.}
\label{}
\end{figure}
%%%%%%%%%%%%%%%%%%% Fig :7 %%%%%%%%%%%%%%%%%%%%%%%%%%%
\begin{figure}[htp]
\centering
\includegraphics[scale=0.20]{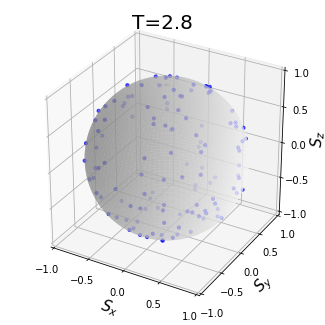}
\includegraphics[scale=0.20]{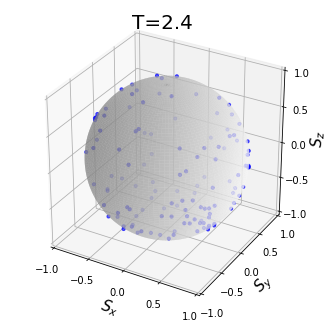}
\includegraphics[scale=0.20]{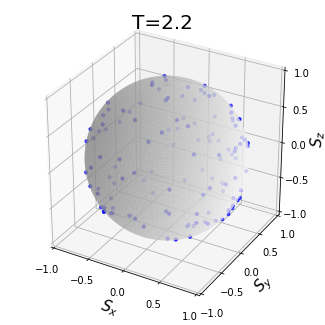}
\includegraphics[scale=0.20]{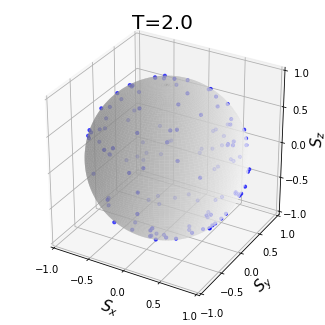}
\includegraphics[scale=0.20]{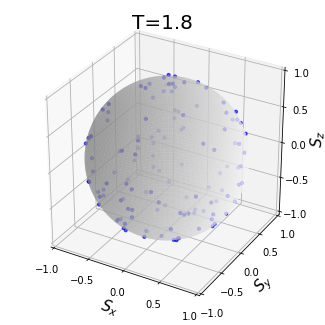}\\
\includegraphics[scale=0.20]{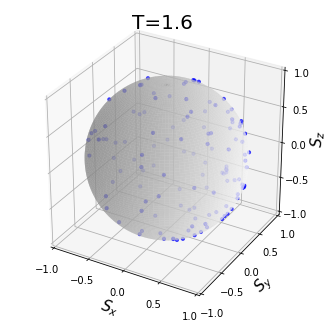}
\includegraphics[scale=0.20]{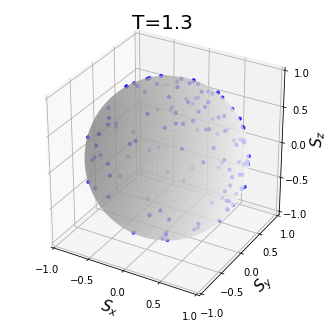}
\includegraphics[scale=0.20]{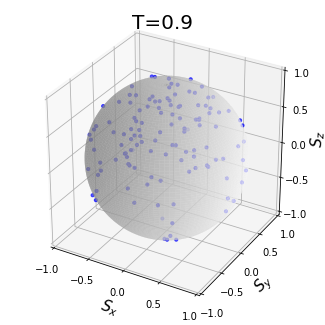}
\includegraphics[scale=0.20]{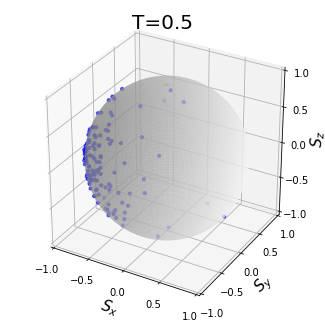}
\includegraphics[scale=0.20]{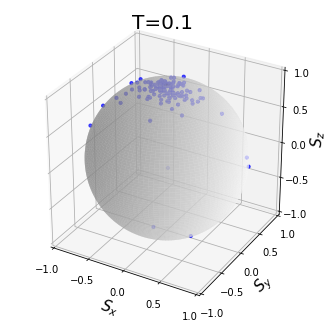}
\caption{Here we are observing the spin configuration at different T for p=0.8.Here we have used lattice size $5 \times 5 \times 5 $.}
\label{}
\end{figure}
%%%%%%%%%%%%%%%%%%%%%%%%%%%%%%%%%%%%%%%%%%%%%%%
After that we have simulate the critical behaviour of {$\bf {A_pB_{1-p}}$} magnetic alloy with p=0.9.So here $10 \% $ atom is B-type which are interacting with each other antiferromagnetically and interacting ferromagnetically with A-type atom.Here we have observed that the system showing almost ferromagnetic behaviour like pure 3D Heisenberg case.Here ferromagnetic ground state is dominating at low temperature,which is expected but $T_c$ is decreased than the previous case.Here we follow the same procedure as before and calculated different thermodynamic variables with T.Here we have same formula for $C_v$ and $\chi$ as before.Here we have presented our results at Fig(4-5).Next
%%%%%%%%%%%%%%%%%%% Fig:8 %%%%%%%%%%%%%%%%%%%%%
\begin{figure}[htp]
\centering
\includegraphics[scale=0.30]{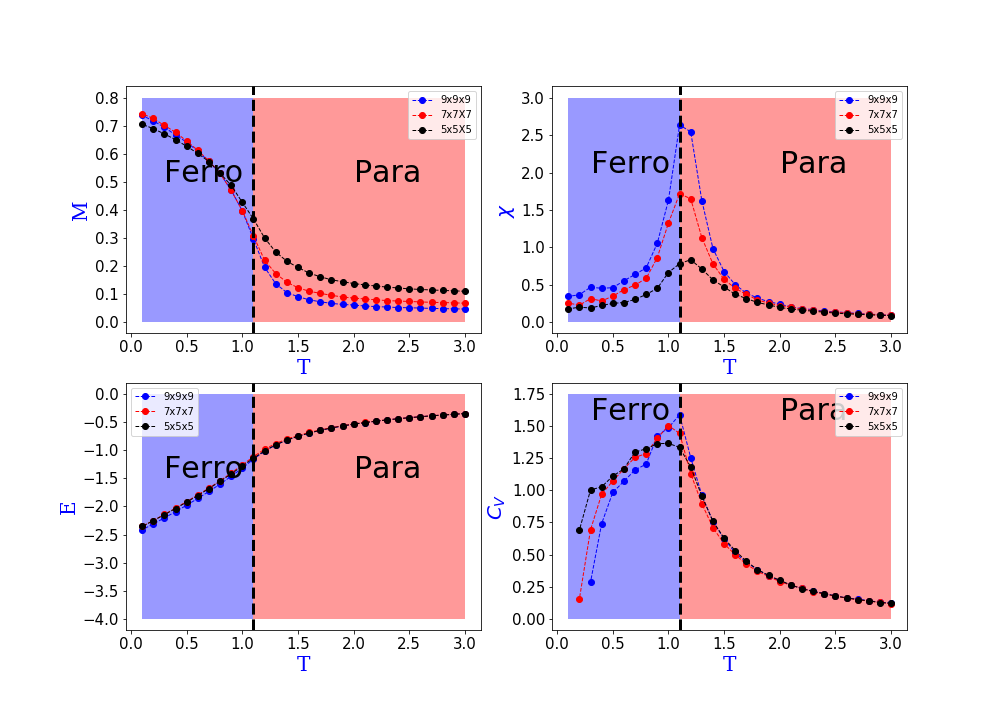}
\caption{Here we have presented the critical behaviour for p=0.60.Here we have presented M,{\large{$\chi$}},E,$C_v$ with T.Here we have presented the critical behaviour of {$\bf{A_pB_{1-p}}$} magnetic alloy.Here A is ferromagnetic atom and B is antiferromagnetic atom.}
\label{}
\end{figure}
%%%%%%%%%%%%%%%%% Fig:9 %%%%%%%%%%%%%%%%%%%%%%%%
\begin{figure}[htp]
\centering
\includegraphics[scale=0.20]{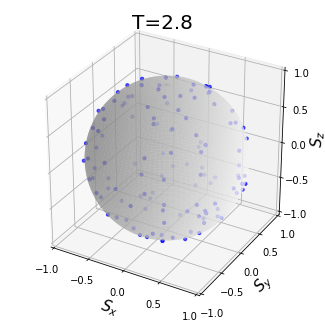}
\includegraphics[scale=0.20]{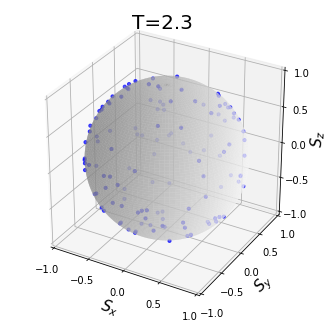}
\includegraphics[scale=0.20]{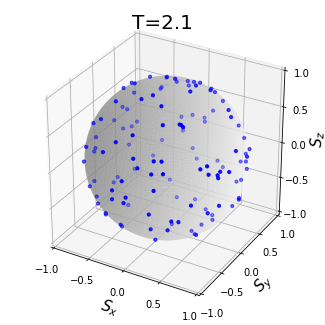}
\includegraphics[scale=0.20]{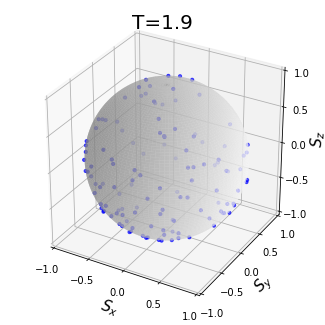}
\includegraphics[scale=0.20]{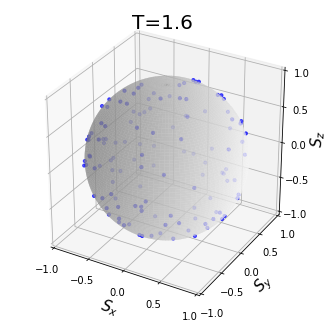}\\
\includegraphics[scale=0.20]{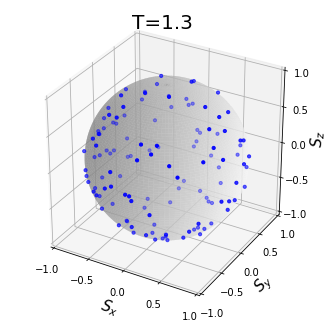}
\includegraphics[scale=0.20]{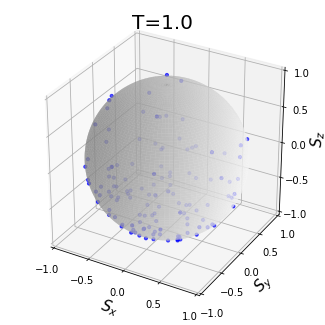}
\includegraphics[scale=0.20]{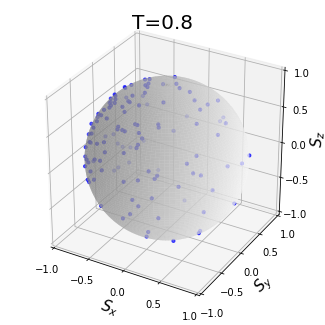}
\includegraphics[scale=0.20]{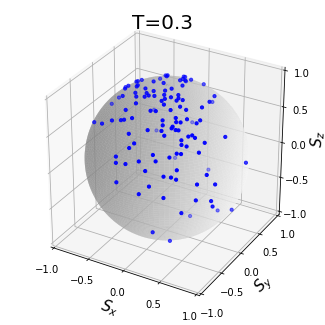}
\includegraphics[scale=0.20]{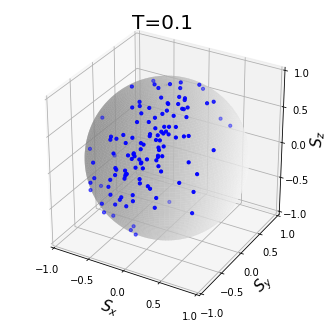}
\caption{Here we are observing the spin configuration at different T for p=0.6.Here we have used lattice size $5 \times 5 \times 5$.}
\label{}
\end{figure}
%%%%%%%%%%%%%%%%%%%%%%%%%%%%%%%%%%%%%%%%%%%%%%%%%
we have simulated the critical behaviour for p=0.8 and for this value of p approximate $20 \% $ atom is B-type.Here we observe that the saturation value of M is decreased in presence of B atom.Here $T_c$ is decreased and here we have used the same
procedure as we mentioned before for calculating the different thermodynamic quantity.Here we have used same formula for calculating M,$C_v$ and $\chi$ Fig(6-7).
%%%%%%%%%%%%%%%%%%%%%%%%%%%%%%%%%%%%%%%%%%%%%%%%%%%
Here we have simulated this phase transition for $5 \times 5 \times 5 $, $7 \times 7 \times 7 $, $9 \times 9 \times 9$ lattice system.Here we observed interesting behaviour of M during phase transition.
%%%%%%%%%%%%%%%%%%%% Fig:10 %%%%%%%%%%%%%%%%%%%%%%%%  
\begin{figure}[htp]
\centering
\includegraphics[scale=0.30]{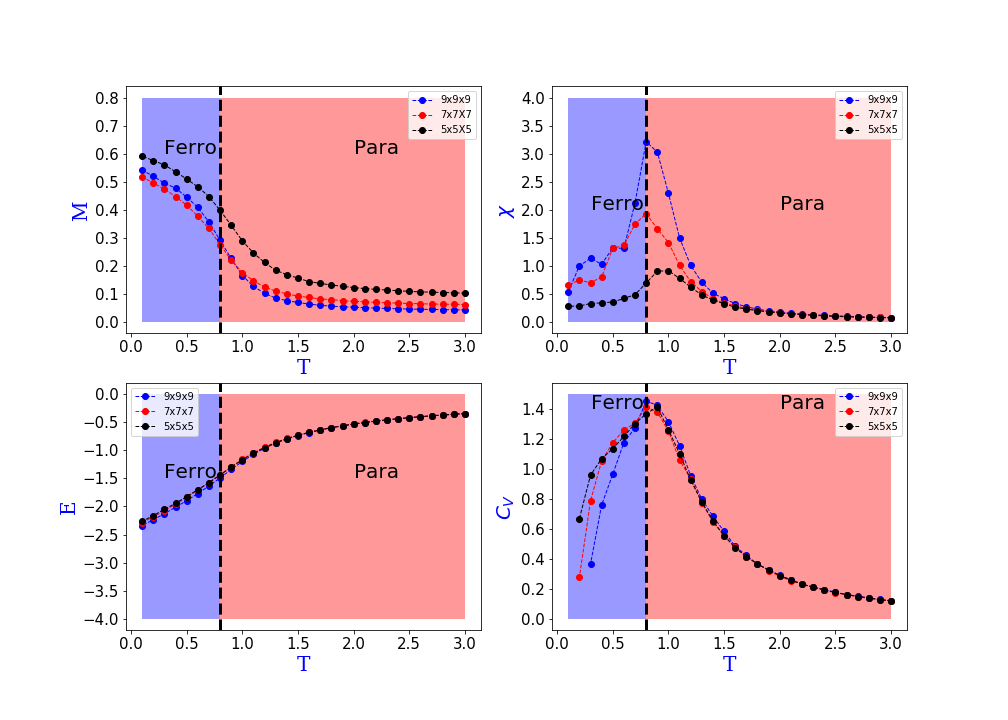}
\caption{Here we have presented the critical behaviour for p=0.50.Here we have presented M,{\large{$\chi$}},E,$C_v$ with T.Here we have presented the critical behaviour of {$\bf {A_pB_{1-p}}$} magnetic alloy.Here A is ferromagnetic atom and B is antiferromagnetic atom.}
\label{}
\end{figure}
Here we have simulated the critical phenomena for p=0.6 Fig(8-9).Here nearly 40\% B atoms is present and from the saturation value of M we can easily observed that the value of M is
%%%%%%%%%%%%%%%%%%%%%%%%%%%%%%%%%%%%%%%%%%%%%%%%%%
decreased to around 0.8.So we can say that in this case a stagger antiferromagnetic order is developed and that is the reason of decreasing the value of saturation value of M.Here $T_c$ is 1.01.Here we have followed the same steps to calculate M ,$C_v$ and $\chi$.Here we have simulated this transition for three different lattice size ($5 \times 5 \times 5$),($7 \times 7 \times 7$) and ($9 \times 9 \times 9$).
%%%%%%%%%%%%%%%%% Fig:11 %%%%%%%%%%%%%%%%%%%%%%%%%%
\begin{figure}[htp]
\centering
\includegraphics[scale=0.20]{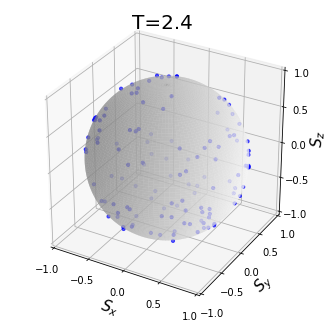}
\includegraphics[scale=0.20]{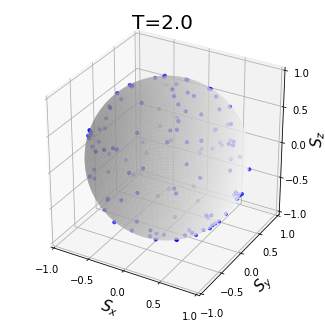}
\includegraphics[scale=0.20]{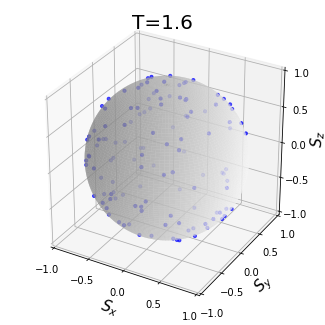}
\includegraphics[scale=0.20]{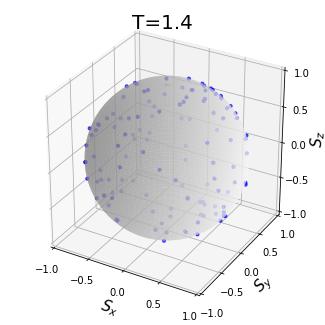}
\includegraphics[scale=0.20]{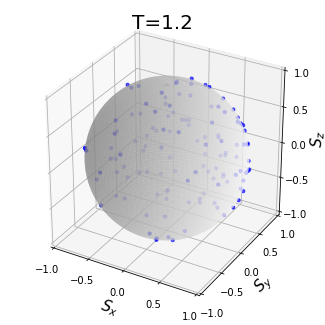}\\
\includegraphics[scale=0.20]{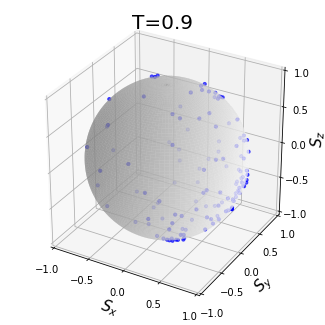}
\includegraphics[scale=0.20]{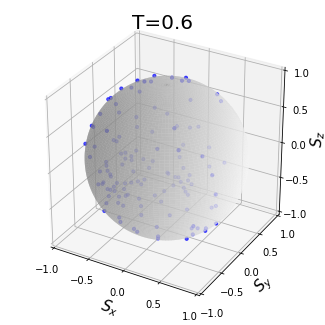}
\includegraphics[scale=0.20]{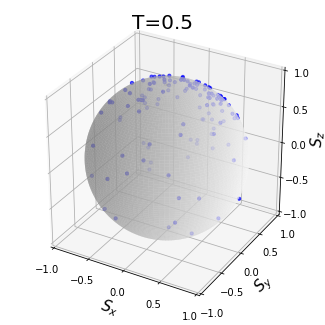}
\includegraphics[scale=0.20]{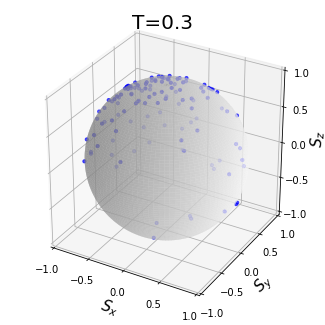}
\includegraphics[scale=0.20]{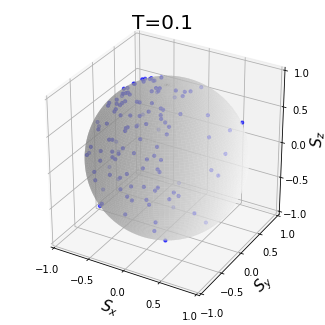}
\caption{Here we are observing the spin configuration at different T for p=0.5.Here we have used $5 \times 5 \times 5$ lattice size.}
\label{}
\end{figure}
Here we have simulated  the critical phenomena for p=0.5 where 50\% atoms are A type and 50\% are B type atoms Fig(10-11).Here we have clearly observed that saturation value of M which indicate the FM ordering is decreasing to 0.5 and clearly indicate the presence of stagger antiferromagnetic ordering.Here we have noticed that $T_c$ is around 0.8.Here we have simulated this critical behaviour for different lattice size which are $5 \times 5 \times 5 $,$7 \times 7 \times 7$ and $9 \times 9 \times 9$.Here we use same procedure of measuring M,$C_v$ and $\chi$ at different T.Here actually we observed the mixed phase and tetracritical behaviour.
Here we have used RG theory for giving a details description of  tetracritical behaviour of that 
dilute magnetic system.Here we use the most popular RG theory based on $\epsilon$ expansion and that RG analysis is used to explain the phase behaviour for two competitive field\cite{nagaosa}.In our case the field components are $n_1=3$ and $n_2=3$.Here four fixed points exist for decoupled case when w=0 and those fixed points are G-G,G-H,H-G and H-H fixed points.Among these fixed points only H-H fixed point is stable fixed point and all RG flowlines are converging to that fixed point Fig(12-Left).That fixed point will determine the critical behaviour and critical exponents of the phase transition and system going to shows tetracritical behaviour and mixed phase.Here the value of all type of fixed points is given in below table.
\begin{figure}[htp]
\centering
\includegraphics[scale=0.40]{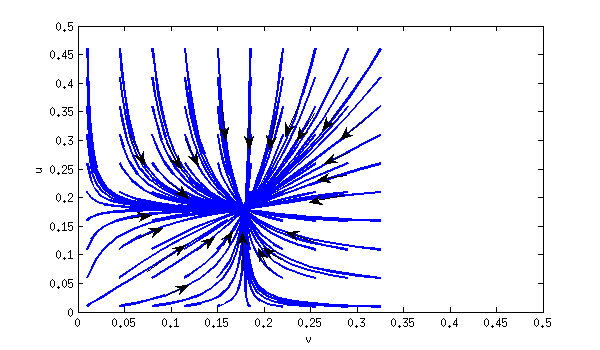}
\includegraphics[scale=0.40]{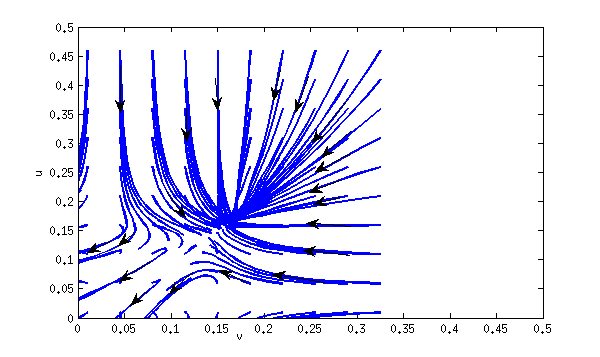}
\caption{\textbf{Left)} RG flowline for w=0.0.Here all flow line are moving to H-H fixed point.\textbf{Right)} RG flow line for w=0.13.Here flow lines are plotted in u-v parameter space.}
\label{}
\end{figure}
\begin{center}
 \bf {TABLE-I}\\
\
\\
\begin{tabular}{|l|l|l|l|}
\hline
	Fixed point & $u^*$ & $v^*$ & $w^*$\\
\hline
	G-G & 0 & 0 & 0\\
\hline
	H-G & $\frac{2{\pi}^2 \epsilon} {(n_1+8)}$ & 0 & 0\\
\hline
	G-H & 0 & $\frac{2{\pi}^2 \epsilon}{(n_2+8)}$  & 0\\
\hline
H-H & $\frac{2{\pi}^2 \epsilon} {(n_1+8)}$ & $\frac{2{\pi}^2 \epsilon}{(n_2+8)}$ & 0 \\ 
\hline   
\end{tabular}
\end{center}
In case of $w \neq 0$ system also have stable fixed point and here biconical fixed point is the most stable one since $n=n_1 +n_2=6 > n_c$,where $n_c=4-2\epsilon +O({\epsilon}^2)$.In this case there exist a "seperatrix" $F(u,v,w) \sim uv-w^2=0$ and above that seperatrix when F(u,v,w) > 0 all the RG flow lines are converging to biconical fixed point and that determine the critical behaviour of that phase transition.Below that seperatrix all the flow lines are showing run-away behaviour.Here biconical fixed point indicate  tetracritical behaviour and mixed phase is most stable phase in this case. 
%%%%%%%%%%%%%%%%%%%%%%%%%%%%%%%%%%%%%%%%%%%%%%%%%%%%%%%%%%%%%%%%%%%%%%%%%%%%%%%%%%%%%
%%%%%%%%%%%%%%%%%%%%%%%%%%%%%%%%%%%%%%%%%%%%%%%%%%%%%%%%%%%%%%%%%%%%%%%%%%%%%%%%%%%%%
%                Discussion and Conclusion                                          %
%                                                                                   %
%%%%%%%%%%%%%%%%%%%%%%%%%%%%%%%%%%%%%%%%%%%%%%%%%%%%%%%%%%%%%%%%%%%%%%%%%%%%%%%%%%%%%
%%%%%%%%%%%%%%%%%%%%%%%%%%%%%%%%%%%%%%%%%%%%%%%%%%%%%%%%%%%%%%%%%%%%%%%%%%%%%%%%%%%%%
\section{Discussion and conclusion}
Here we have simulated successfully the critical behaviour at different value of doping(p) for dilute magnetic alloy {$\bf{A_pB_{1-p}}$}.At the intermediate value of doping we are observing that two independent second order phase transition is taking place.Here we have only studied the Ferro-para transition and observe that the saturation value of spontaneous magnetization(M) is reaching  near at 0.5.That clearly says that the rest of the magnetization is cancelled out in presence of stagger antiferromagnetic ordering.So here two independent phase transition are taking place and at low temperature two competitive order FM and AFM coexisting.So this is the sign of tetracritical behaviour where two second order phase transition takes place.Here we able to probe the mixed phase behaviour successfully and observe the spin configuration of that mixed phase.In fact in this work we have simulated the effect of disorder,impurity,doping in a 3D Heisenberg model and study the critical behaviour.We have noticed that when doping strength is low(p=0.9) then it is not effecting too much the nature of phase transition.Here the doping of antiferromagnetic atom \textbf{B} is working as magnetic impurity or disorder when the doping strength is low.The present simulation results find good agreement with previously reported results in this topic\cite{bekhechi2004chiral}\cite{beath}. 

%%%%%%%%%%%%%%%%%%%%%%%%%%%%%%%%%%%%%%%%%%%%%%%%%%%%%%%%%%%%%%%%%%%%%%%%%%%%%%%%%%%%
%              Acknowledgement                                                     %
%%%%%%%%%%%%%%%%%%%%%%%%%%%%%%%%%%%%%%%%%%%%%%%%%%%%%%%%%%%%%%%%%%%%%%%%%%%%%%%%%%%%
\section{Acknowledgement}
N.B  is greatly acknowledging University of Seoul(UOS) for the funding from SAMSUNG and NRF project at initial stage of this work and we are greatly acknowledging IIT kanpur for giving visiting scholar position and providing generous research facility during the visit.Author is acknowledging Prof.P.K.Mukherjee,Prof.M.Acharyya, Prof.T.Das,Prof.J.Jung for several discussion about different aspect of the phase transition and critical phenomena over the years.Author is thankful to Prof.Chandan Dasgupta and Prof.Sumilan Banerjee for several lecture on disorder system during the course at IISc,Bangalore.Also author is thankful to Prof.Thomas Vojta,Prof.Altland,Prof.F.Evers,Prof.Dietrich Belitz for several interesting lecture on phase transition during the conference at IMSc,Chennai.Author is greatly thankful to Prof.Allan H MacDonald for his fruitful collaboration with our group and several help during the progress.Author is thankful to Prof.Aharony for notifying one of his seminal paper in this direction. 

%%%%%%%%%%%%%%%%%%%%%%%%%%%%%%%%%%%%%%%%%%%%%%%%%%%%%%%%%%%%%%%%%%%%%%%%%%%%%%%%%%%%
%              Reference                                                           % 
%%%%%%%%%%%%%%%%%%%%%%%%%%%%%%%%%%%%%%%%%%%%%%%%%%%%%%%%%%%%%%%%%%%%%%%%%%%%%%%%%%%%

\end{document}